# Indentation-induced polar domain reconfiguration in the charge-density-wave semiconductor $EuTe_4$

Hidefumi Takahashi[1,2*], Yuki Umewaka[1], Kiichiro Yoshida[1], Ritsuto Nakamura[1], Akitoshi Nakano[3], Takao Matsumura[1], Daiki Sekine[4#], Masakazu Matsubara[1,2,5], Masayuki Ochi[6,7,8], and Shintaro Ishiwata[1,2†]

[1]*Division of Materials Physics and Center for Spintronics Research Network, Graduate School of Engineering Science, The University of Osaka, Osaka 560-8531, Japan*

[2]*Spintronics Research Network Division, Institute for Open and Transdisciplinary Research Initiatives, The University of Osaka, Yamadaoka 2-1, Suita, Osaka, 565-0871, Japan*

[3]*Department of Physics, Nagoya University, Nagoya 464-8602, Japan*

[4]*Department of Physics, Graduate school of science, Tohoku University, Aoba-ku, Sendai 980-8579, Japan*

[5]*PRESTO, Japan Science and Technology Agency (JST), Kawaguchi 332-0012, Japan*

[6]*Department of Physics, The University of Osaka, Machikaneyama-cho, Toyonaka, Osaka 560-0043, Japan*

[7]*Forefront Research Center, The University of Osaka, Machikaneyama-cho, Toyonaka, Osaka 560-0043, Japan*

[8]*RIKEN Center for Computational Science, Saitama, Japan*

**Controlling polar domains in conducting materials by external fields could unlock polarisation-based electronic functionalities, yet remains experimentally challenging. While mechanical actuation driven by local strain, stress, or strain gradients has been proposed as an external-voltage-free route, its experimental demonstration in relatively conducting polar materials remains limited. Here we demonstrate mechanically induced polar-domain reconfiguration in the charge-density-wave (CDW) semiconductor $EuTe_4$ at room temperature, whose resistivity is on the order of 10 mΩcm. Polarised-light microscopy and second-harmonic-generation measurements reveal CDW-induced in-plane polarisation domains with an approximately 90° domain configuration. Local indentation with a ~1 μm-apex-radius needle reconfigures these domains, producing write/erase-like changes in wedge-shaped regions. These results establish local mechanical-field domain engineering in a relatively conducting CDW semiconductor, with possible contributions from both ferroelastic-like and flexoelectric mechanisms.**

Hidefumi Takahashi (corresponding author)
*takahashi@mp.es.osaka-u.ac.jp or takahashi.hidefumi.es@osaka-u.ac.jp
†ishiwata.shintaro.es@osaka-u.ac.jp

## INTRODUCTION

Precise control of polar domains is essential for low-power memories and multifunctional devices[1,2]. Conventional electric-field-driven switching, however, requires high electric fields, restricting practical implementation to high-resistivity materials in order to suppress leakage currents[3–5]. Recent studies have demonstrated that a mechanically loaded scanning-probe tip can manipulate polarisation without an applied voltage.[6–12] Unlike uniform strain, a highly localised force applied by a scanning-probe tip generates both anisotropic stress/strain fields and strain-gradient fields. The former can favour particular ferroelastic-like or structural variants by modifying their relative elastic energies, whereas the latter can act as an effective flexoelectric bias coupled to the polar distortion.[6,9,13,14] Thus, local mechanical stimulation can reconfigure polar or polar-structural domains through ferroelastic-like variant selection, strain-gradient-induced flexoelectric coupling, or a combination of both. In $BiFeO_3$ thin films, for example, the scan direction can select domain-rotation pathways, realizing mechanically driven multiaxial domain control.[6] Similar tip-

# Present address: Quantum Materials and Applications Research Center, National Institutions for Quantum Science and Technology, Takasaki, Gunma 370-1292, Japan

force-induced manipulation has also been reported in the van der Waals ferroelectric $CuInP_2S_6$, where large local deformation and strain gradients enable room-temperature domain switching.[15,16] More broadly, van der Waals materials can accommodate large elastic deformation and reduced substrate clamping, providing a favourable platform for nanoscale mechanical domain engineering. [17]

The mechanical control of polar domains has largely been studied in wide-band-gap, low-conductivity materials. A key outstanding question is whether mechanically induced domain reconfiguration can also be achieved in highly conducting polar metals and semimetals,[18–21] and in narrow-gap polar semiconductors.[17,22–24]. Recent theoretical and experimental studies have shown that strain-gradient-induced flexoelectric coupling can induce or control polar distortions even in metallic or conducting systems[25–27]. In previous experimental demonstrations, however, built-in strain-gradient fields associated with interfaces, epitaxial constraints, or twisted/moiré structures have played central roles, rather than externally applied local mechanical stimuli. Extending external mechanical control to intrinsic polar domains in conducting crystals could open opportunities in several areas: in spintronics, where current-driven magnetic textures could be combined with electric- or strain-mediated switching [28–30]; in topological materials, where Berry-curvature sources could be mechanically reconfigured through domain engineering[31,32]; in thermoelectrics, where phonon-scattering centres could be tailored without changing carrier density[33].

Here, we focus on the external mechanical reconfiguration of intrinsic CDW-induced polar domains in bulk $EuTe_4$, a van der Waals material with a narrow band gap (<0.2 eV)[34–37]. $EuTe_4$ hosts a CDW state that breaks inversion symmetry and produces an in-plane polar structure. Despite its gapped Fermi surface, the room-temperature resistivity remains on the order of 10 mΩcm with a carrier density of $\sim 10^{19}$ $cm^{-3}$.[38] Using polarised-light microscopy, birefringence microscopy, and optical second-harmonic-generation measurements, we image these polar structural domains and show that localised pressure applied by a micrometer-scale needle tip reconfigures the CDW-induced polar domains. These results provide a route toward strain- and stress-based control of CDW-coupled polar/structural domains in a relatively conducting semiconductor.

## RESULTS AND DISCUSSION

At room temperature, $EuTe_4$ stabilizes a CDW state (Fig. 1a). The lattice parameters of the CDW state are tripled along the *b* axis and doubled along the *c* axis relative to the putative non-CDW state (see Fig. S1 in Supplementary Information). Here we denote this state as "primitive CDW state" as the actual structure has been found to exhibit higher-order modulations. Electrical transport measurements show a pronounced hysteresis near room temperature, indicative of a thermal evolution between distinct high- and low-temperature CDW phases (Fig. 1b)[39]. In the high-temperature phase (CDW-I), X-ray diffraction reveals incommensurate superlattice reflections characterised by modulation vector $\boldsymbol{q} \sim 0.67\boldsymbol{b}^*$[39,40]. Upon cooling, additional reflections emerge that are attributable to a low-temperature CDW phase (CDW-II)[38]. High-resolution synchrotron X-ray diffraction further suggests that the large hysteresis arises from a moiré-like superstructure produced by the coexistence of two incommensurate modulations with $\boldsymbol{q} \sim 0.67\boldsymbol{b}^*$[41,42]. In both phases, a subset of Te sites within the two-dimensional Te square-net layers displaces to form trimers (Fig. 1a), breaking in-plane inversion symmetry and generating an in-plane polar structure. The accompanying reduction of crystallographic symmetry from tetragonal to orthorhombic, with inequivalent *a* and *b* axes, naturally leads to the formation of orthorhombic domains. Although CDW-I and CDW-II exhibit incommensurate modulations, their essential structural motif—the Te-trimer formation and the associated inversion-symmetry breaking—is well captured by the primitive CDW state.

In the following, we adopt this primitive CDW description as the basis for discussing the mechanically induced reconfiguration of CDW-coupled polar/structural domains.

Figure 1c shows a polarised-light micrograph of a $EuTe_4$ single crystal, revealing two domains with sharply bifurcated contrast. The domains are related by an approximate 90° rotation of the crystallographic *a* and *b* axes. Complementary two-dimensional birefringence mapping in reflection (Fig. 1d) visualises the optical principal axis as a vector field arising from the in-plane anisotropy. This map resolves two domains whose principal-axis orientations differ by ~90°, consistent with the polarised-light image and with orthorhombic twinning. These observations establish that $EuTe_4$ forms well-defined domains within the *ab* plane as a consequence of its orthorhombic symmetry. Moreover, the crystal further exhibits an in-plane polar structure with broken inversion symmetry; in the insulating limit, this axis would host a spontaneous polarisation.

To clarify the orientation of the polarisation, we performed SHG measurements in a reflection geometry (Fig. 2a). Since SHG is extremely sensitive to inversion-symmetry breaking[43], it enables the detection of polarisation even in highly conductive polar materials[44]. For the stripe-like domains (Fig. 2d), we recorded the angular dependence of the SHG intensity (Fig. 2c) and used the maxima to extract the in-plane polarisation vector. Figure 2b presents a two-dimensional SHG intensity map acquired with the incident fundamental (ω) and emitted second-harmonic (2ω) polarisations set parallel, revealing contrast modulation along the stripe direction. The corresponding polarisation vectors (yellow arrows in Fig. 2b) are oriented at approximately 45° with respect to the stripe direction and rotate by about 90° between adjacent domains. These features are characteristic of the 90° polar domain topology commonly observed in in-plane–polarised two-dimensional ferroelectrics with tetragonal or pseudo-tetragonal symmetry[17,45], indicating that an analogous domain structure is realised in $EuTe_4$ (see Fig. S2 in Supplementary Information for a schematic of the polarisation and domain configuration). Note that although $EuTe_4$ is orthorhombic, the *a* and *b* lattice parameters are sufficiently close that it can be regarded as pseudo-tetragonal for describing the in-plane polar domain topology.

From the above results, we conclude that $EuTe_4$ hosts a domain structure with an in-plane polarisation vector. This, in turn, suggests that local mechanical fields could be used to reconfigure the polar-domain orientation. To test this possibility, we applied localised strain by gently indenting the crystal surface with a needle of ~1 μm apex radius and monitored the ensuing domain evolution using polarised-light microscopy. Preliminary tests using needles with different apex radii showed that the ~1 μm-apex-radius needle was most suitable for inducing wedge-shaped birefringence contrast under the present manual indentation conditions (Supplementary Fig. S3). The experimental arrangement is shown in Figs. 3a,b: the crystal was affixed to a glass slide with a fluorinated adhesive (CYTOP, AGC), the tip was brought into contact from above, and the process was recorded *in situ* (see Supplementary Video 1). We also confirmed that similar mechanically induced birefringence contrast changes were observed when $EuTe_4$ crystals were mounted on a rigid copper plate or a softer Kapton tape, indicating that the response is not governed solely by the macroscopic substrate stiffness (Supplementary Fig. S4). Following indentation (Fig. 3c), a wedge-shaped region nucleates from the contact point with optical contrast distinct from the surrounding matrix; its mean width is ~1 μm and its mean length is ~7 μm, as determined from the statistical analysis of 29 wedge-shaped regions described in Supplementary Fig. S5. The observed contrast indicates a large change in the local optical-axis orientation between adjacent domains, consistent with strain-induced reconfiguration of the orthorhombic (pseudo-tetragonal) CDW-coupled polar/structural domains. To examine whether this contrast arises from irreversible surface damage, we performed SEM observations on the mechanically modified regions. No clear cracks, scratches, or permanent surface deformation were observed in most wedge-shaped reconfigured regions (see Supplementary Fig. S6), indicating that the modified optical contrast is not simply a topographic artifact caused by macroscopic surface deformation.

To examine the write/erase-like response, we applied local strain near the previously written wedge-shaped domain and tracked the subsequent change in birefringence contrast using polarised-light microscopy (see Supplementary Video 2 for the indentation sequence). Figure 3d shows the image after the second indentation. Focusing on the domain marked by the yellow arrow in Fig. 3c, we find that this domain contrast is suppressed upon the second indentation, while a new domain (green arrow) appears in its vicinity. The write/erase-like contrast change occurred on the timescale of the manual mechanical operation, typically on the order of seconds, which should be regarded as an upper bound limited by the manipulation and imaging conditions rather than the intrinsic domain dynamics. These observations indicate that local mechanical stimulation can induce, suppress, or redistribute CDW-induced polar-domain contrast. The selective suppression of the domain contrast near the second indentation site, while more distant domains remain largely unchanged, likely reflects the highly nonuniform mechanical fields generated by local indentation, which decay away from the contact region. This write/erase-like response highlights the potential of local mechanical stimulation for modifying CDW-induced polar-domain states in $EuTe_4$.

Figure 4a shows the birefringence map acquired in the same field of view as Fig. 3d. A clear rotation of the optical principal axis, shown by the colour plot of the orientation angle φ, within the wedge-shaped region is observed relative to the surrounding matrix. An enlarged view of the area exhibiting the largest domain change (black box) is presented in Fig. 4b. To quantify the spatial evolution of the principal-axis orientation within the wedge domain, Fig. 4c plots line scans along the three trajectories (I–III) indicated in Fig. 4b. In particular, along line I, the principal-axis angle varies strongly inside the wedge, and the magnitude of this variation decays gradually with distance from the indentation point. The maximum change in φ reaches approximately 60°, indicating a substantial local modification of the optical anisotropy. Because the optical anisotropy in $EuTe_4$ is associated with the CDW-coupled polar/structural distortion, this large local change strongly suggests that the axis orientation of the CDW-coupled polar/structural domain is modified by local mechanical strain. Optical-axis analysis over a wider field of view, including regions outside the wedge-shaped feature, also shows that the wedge-shaped region exhibits a significantly larger φ change than nearby deformation-related contrast features (Supplementary Fig. S7). SHG microscopy after indentation shows a clear polarisation-dependent contrast in the wedge-like region (Supplementary Fig. S8). When both the incident fundamental light and detected SHG light are polarised parallel to the crystallographic *b* axis, the wedge-like region is relatively bright compared with the surrounding matrix. In contrast, when both polarisations are parallel to the crystallographic *a* axis, the contrast is reversed. This reversal of the SHG contrast indicates that the nonlinear optical response associated with the polar structure is modified in the mechanically reconfigured region, supporting reconfiguration of CDW-coupled polar/structural domains.

To further examine whether the mechanically induced birefringence patterns are related to the local mechanical-field geometry, we performed additional experiments with different needle orientations relative to the crystallographic axes, different needle inclination angles relative to the *ab* plane, and mechanical scanning along different in-plane directions. Representative results are shown in Fig. 5, with more detailed images provided in Supplementary Figs. S9 and S10. The resulting wedge-like or line-like birefringence contrast was not random, but depended on the needle orientation, contact geometry, and scanning direction. When the needle direction was varied with respect to the *a* and *b* axes, the reconfigured regions consistently appeared along directions approximately 45° from the crystallographic *a* and *b* axes, indicating that the mechanically induced reconfiguration is strongly constrained by the crystallographic anisotropy of the CDW-coupled polar/structural domains. Changing the needle inclination angle α also modified the spatial extent of the contrast features: at smaller inclination angles relative to the *ab* plane, the line-like or wedge-like domains tended to extend over longer distances. In addition, when the needle was mechanically scanned approximately along the *a* axis, wedge-like domains preferentially formed in the forward direction of the needle motion, whereas scanning approximately along the *b* axis produced less well-defined and more spatially broad contrast changes near the scanned regions. These crystallographic and scanning-direction dependences provide evidence that the mechanically induced domain reconfiguration is consistent with possible contributions from local anisotropic strain/stress fields and local strain-gradient fields determined by the contact geometry and needle motion. We also examined the effect of CDW thermal history by performing indentation experiments after heating the sample to 403 K and after cooling it to 81 K, followed by measurements at room temperature (Supplementary Fig. S11). Similar line-like or wedge-like birefringence contrast changes were observed in both cases, suggesting that the mechanically induced reconfiguration can occur in both CDW-I- and CDW-II-related thermal histories and is primarily associated with the in-plane Te-trimer distortion.

We now discuss the correspondence between the observed domain behaviour and the crystal structure of $EuTe_4$. As shown in Fig. 1a, the two-dimensional Te layers form a square net; within each unit (red box), the central Te atom is displaced by ~0.2 Å along the *a* axis, forming a trimer. This displacement breaks in-plane inversion symmetry and stabilises an orthorhombic CDW state, defining an in-plane polarity oriented along the *a* axis. The observed 90° domains can be understood as neighboring CDW-coupled polar/structural variants whose Te-trimer displacement directions are rotated by 90° with respect to each other. In this picture, the domain wall is expected to run approximately 45° from the Te-displacement, or polar-distortion, direction, as schematically shown in Fig. 6a. This is consistent with the domain-wall orientations observed in polarised-light microscopy and with the in-plane polar-axis directions inferred from SHG measurements. We therefore attribute the 90° domain structure in $EuTe_4$ to the coexistence of regions with orthogonal Te-displacement directions in the square-net Te layer.

Figure 6b schematically summarizes the possible relationship between Te displacement, CDW stability, and local mechanical deformation. In the putative non-CDW state with tetragonal symmetry, the Te atom resides at the centre of the square lattice. In the CDW state, by contrast, Te atoms are displaced in-plane to form Te trimers, producing a CDW-induced polar structural distortion. First-principles calculations yield a relative energy difference (enthalpy difference) of $\Delta H \equiv H_{CDW} - H_{non} \sim -0.0164$ eV/f.u. in the absence of strain, indicating that the CDW structure with Te trimers is energetically favoured over the undistorted phase. We emphasize that $\Delta H$ represents the relative stability of the CDW and non-CDW structures and should not be regarded as an activation barrier for switching between two 90°-rotated polar-domain variants. In the absence of local strain, the two in-plane Te-displacement directions are crystallographically equivalent and energetically degenerate, as schematically shown in the left panel of Fig. 6b. When local strain is applied using a sharp tip, however, the induced strain field can be highly anisotropic. In the CDW phase of $EuTe_4$, the Te displacement along the *a* axis makes the crystallographic *a* and *b* axes inequivalent, with the *a* axis being shorter than the *b* axis ($3a < b$). Therefore, anisotropic strain or stress is expected to modify the relative stability of CDW-coupled polar/structural variants. To examine this effect, we performed first-principles calculations under two types of anisotropic strain, effectively corresponding to tensile or compressive deformation along the *a* axis. The calculated results are summarized in Table 1. Tensile strain along the *a* axis makes $\Delta H$ more negative, changing it from −0.0164 to −0.0175 eV/f.u., indicating stabilization of the CDW state. In contrast, compressive strain along the *a* axis, equivalently tensile strain along the *b* axis, reduces the magnitude of $\Delta H$ to −0.0145 eV/f.u., indicating destabilization of the CDW state relative to the non-CDW structure. These results suggest that local anisotropic strain or stress can modify the relative stability of CDW-induced polar/structural states, as schematically shown in the middle panel of Fig. 6b. This strain sensitivity provides a possible microscopic basis for mechanically induced domain reconfiguration in $EuTe_4$. Local indentation by a sharp tip is expected to generate highly nonuniform mechanical fields, including both anisotropic stress/strain and strain gradients. Therefore, the observed reconfiguration of CDW-induced polar-domain contrast may be understood in terms of at least two possible, and not mutually exclusive, mechanisms.

One possible contribution is anisotropic-stress-induced ferroelastic-like reconfiguration of CDW-coupled structural domains. In quasi-two-dimensional rare-earth tritellurides, in-plane uniaxial stress has been shown to realign the CDW ordering direction, demonstrating that CDW configurations can be strongly controlled by anisotropic strain [46]. This is consistent with our first-principles calculations showing that the relative stability of the CDW structure in $EuTe_4$ is sensitive to anisotropic strain. Thus, local indentation may reconfigure the observed domain contrast by changing the relative stability of CDW-coupled polar/structural variants. At the same time, the estimated stress required to strongly destabilize the CDW state is on the order of several GPa, comparable to pressures that can induce structural damage in van der Waals crystals. Therefore, stress-induced CDW-variant selection may contribute near the tip-contact region, but it may not fully explain non-destructive reconfiguration over a wider micrometer-scale region.

Another possible contribution is a strain-gradient-induced flexoelectric bias. Local indentation can generate a radial and highly nonuniform strain-gradient field, which may produce an effective flexoelectric field near the tip (see Fig. S12 in the Supplementary Information)[6,47]. This field can lower the energy of CDW-induced polar/structural domains whose polar distortion, namely Te displacement, is aligned with the effective flexoelectric field, as schematically shown in the right panel of Fig. 6b. Previous studies on polar materials have shown that mechanically concentrated strain-gradient fields, such as those generated near crack or indentation regions, can induce local polar-domain responses[6,15,47,48]. In the present experiments, wedge-shaped reconfigured regions also emerge from the mechanically affected region near the needle contact, which is qualitatively consistent with a flexoelectric-bias scenario. In this case, the local strain-gradient field may preferentially stabilize CDW-induced polar distortions aligned with the effective flexoelectric field and thereby promote domain reconfiguration.

Thus, our experimental observations are compatible with both a strain-gradient-induced flexoelectric bias and a ferroelastic-like reconfiguration mechanism driven by anisotropic stress or strain. The flexoelectric scenario is supported by the similarity to mechanically generated strain-gradient effects in polar materials, whereas the ferroelastic-like scenario is supported by the known sensitivity of CDW order to uniaxial strain and by our calculated strain dependence of CDW stability. However, the present experiments do not quantitatively determine the local strain-gradient magnitude, flexoelectric field, switching barrier, or stress distribution, and therefore do not uniquely isolate either contribution. Nevertheless, the present results show that local mechanical stimulation can modify CDW-induced polar-domain contrast in the relatively conducting CDW semiconductor $EuTe_4$. This finding extends mechanically induced polar-domain reconfiguration to a relatively conducting CDW semiconductor, beyond the conventional focus on insulating ferroelectrics. Quantitative force-controlled indentation or scanning experiments, combined with finite-element modeling, will be required to reveal the dominant microscopic mechanism and to evaluate possible device functionalities.

## CONCLUSIONS

In conclusion, we demonstrate mechanically induced reconfiguration of CDW-induced polar domains in the relatively conducting narrow-gap semiconductor $EuTe_4$ at room temperature and under ambient conditions. Polarised-light microscopy, two-dimensional birefringence mapping, and SHG imaging reveal in-plane polar structural domains associated with the CDW-induced Te-trimer distortion. Local indentation with a micrometer-scale needle produces wedge-like birefringence contrast, indicating reconfiguration of CDW-coupled polar/structural domains. First-principles calculations show that the CDW-induced polar structure is sensitive to anisotropic strain, providing a possible microscopic basis for the mechanical response. The observed reconfiguration may involve both strain-gradient-induced flexoelectric bias and stress/strain-induced ferroelastic-like reconfiguration of CDW-coupled structural domains. Our results extend mechanically induced polar-domain reconfiguration to a relatively conducting CDW semiconductor and provide a platform for exploring strain- and stress-based control of polar structural domains in materials with diverse electronic conductivities.

## METHODS

### Crystal growth

Single crystals of $EuTe_4$ were grown using the Te flux method[38,39]. Eu lumps (99.9 %) and Te granules (99.9999 %) were mixed in a molar ratio of 1 : 15. The weighed starting materials were sealed in an evacuated fused silica tube, followed by heating at 850 °C for 2 days in a muffle furnace. The furnace was slowly cooled to 430 °C in 24 h, and held at this temperature for two weeks, followed by decanting using a centrifuge. The single crystals were characterised by a single-crystal x-ray diffractometer (Rigaku XtaLAB mini II) and by synchrotron XRD measurements at BL02B1, SPring-8.

### Measurement of domain structure

$EuTe_4$ domain structures were measured at room temperature using polarised optical microscopy (EVIDENT) and a two-dimensional birefringence measurement system (PA-micro, Photonic Lattice). To resolve the direction of the domain polarity (polarisation), we performed optical second harmonic generation (SHG) measurements. The measurements were carried out using light pulses with a central wavelength of 800 nm, a pulse width of <55 fs, and a repetition rate of 10 kHz. The $EuTe_4$ crystal surface was irradiated at an incident angle of approximately 20°, and the reflected SHG signal was recorded using a charge-coupled device (CCD) camera[49]. All observations were conducted at room temperature under ambient conditions.

### Enthalpy calculations

The structural relaxation for enthalpy calculations of $EuTe_4$ was performed within the density functional theory (DFT) using the Perdew-Burke-Ernzerhof (PBE) exchange-correlation functional[50] as implemented in Vienna ab initio simulation package[51–55]. The projector augmented wave (PAW) method[56] was used to account for the treatment of core electrons. We neglected the Eu $f$ orbitals with the open-core treatment since they are predicted to be more than 1.5 eV below the Fermi level[57]. The cut-off energy for plane waves forming the basis set was set to 500 eV. We took 18×18×6 and 18×6×3 $k$-meshes for non-CDW and CDW phases, respectively.

We evaluated the enthalpy $H$ under the stress $\sigma$ using the relation, $H = E_0 - \sum_i \sigma_{ii}\epsilon_{ii} V_0$, where $E_0$, $\epsilon$, $V_0$ are the total energy without stress, the strain, and the volume of the cell without stress, respectively. Here, we assumed that the off-diagonal components of the strain and stress tensors are zero. We optimized the internal coordinates and *c*-axis length with fixed *a*- and *b*-axis lengths until the Hellmann-Feynman force on every atom became less than 1 meV/Å, from which we obtained the relationship between the stress values and the lattice constants (see Note 4 in the Supplementary Information). The initial lattice parameters and atomic positions were taken from the single-crystal X-ray diffraction experiment[38].

**Supplementary information**

Additional data, figures and information are available in the supplemental information.

**ACKNOWLEDGEMENTS**

Not applicable.

**Funding**

This study was supported in part by KAKENHI (Grant No. JP23H04871, JP23K13059, JP24H00413, JP24H01621, JP24H01639, JP24K00570, JP25H00420, JP25H01248, JP25K22081, JP25K08457, and JP26H02014), FOREST (No. JPMJFR236K), PRESTO (Grant No. JPMJPR23H9), and CREST (Grant No. JPMJCR24R5) from JST, Murata Foundation, Yazaki Memorial Foundation for Science, Technology, and Asahi Glass Foundation.

**Author Contributions**

Crystal growth and resistivity measurements were carried out by H.T, K.Y, and R.N. XRD was carried out by A.N. Polarised optical microscopy and birefringence measurements were carried out by H.T and Y.U. SHG measurement was carried out by D.S, T. M. and M.M. Enthalpy calculations were carried out by M.O. This paper was written by H.T. and S.I. with input from other authors. The supplement was written by H.T. and S.I. with input from other authors. H.T. and S.I. supervised the experimental and theoretical part of this work. This project was conceived by H.T.

**Data Availability**

The authors declare that all the data that support the findings of this study are available within the paper and Supplementary Information. Additional relevant data are available from the corresponding authors upon reasonable request. Source data are provided with this paper. Contact Hidefumi Takahashi (takahashi.hidefumi.es@osaka-u.ac.jp)

**Competing Interests**

The authors declare no competing interests.

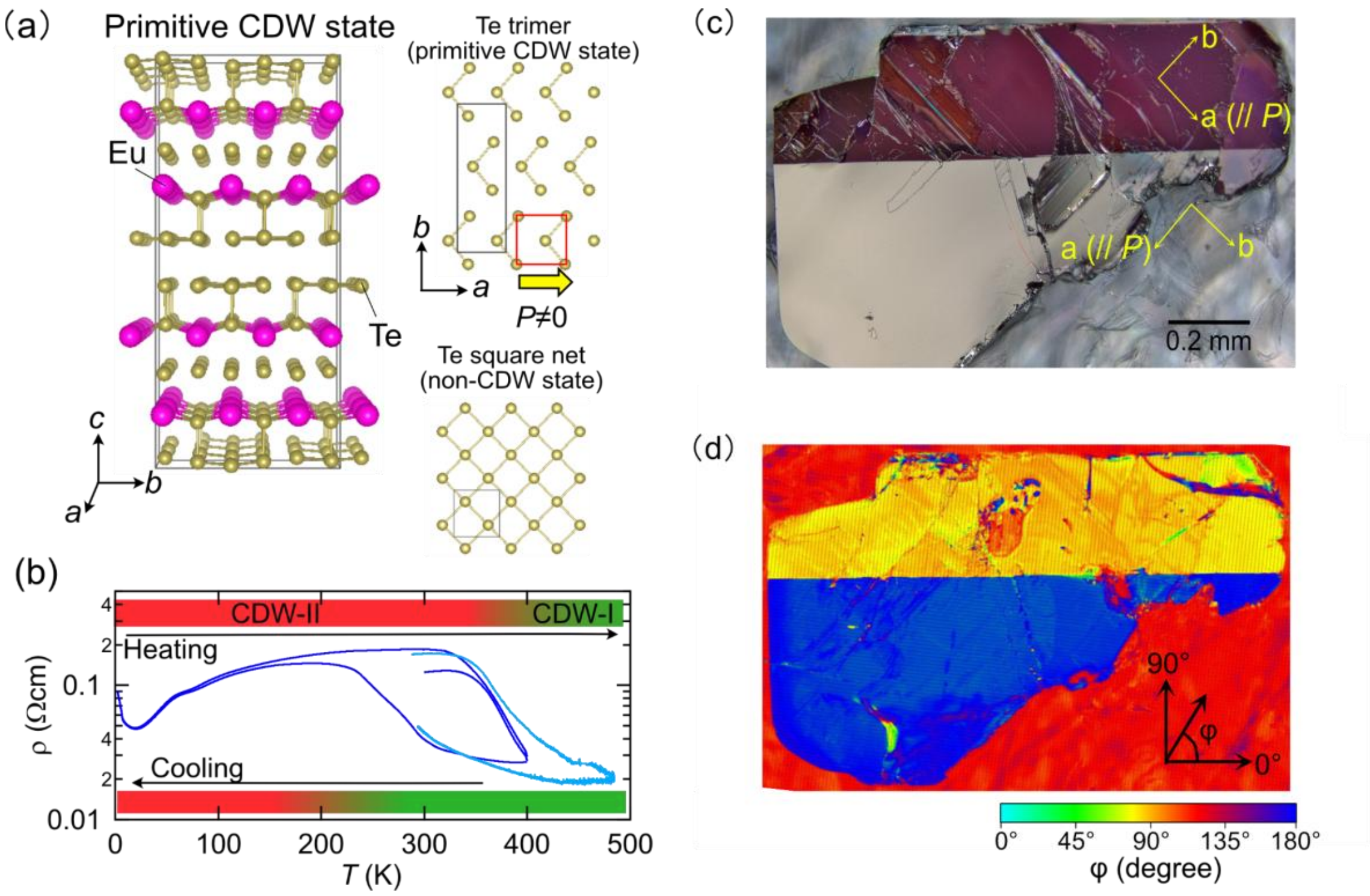

**Fig. 1 Charge-density-wave state in $EuTe_4$.** (a) Crystal structure of $EuTe_4$ in the primitive CDW state[58]. The two-dimensional Te layers are highlighted; the Te layer in the primitive CDW state is shown at top right, and that in the non-CDW state is shown at the bottom right for comparison. (b) Temperature dependence of the in-plane electrical resistivity $\rho(T)$ of $EuTe_4$. The colour bars on top and bottom represent the phase transitions between CDW-I and II on heating and cooling, respectively. (c) Polarised optical micrograph of a single crystal, revealing two domains with distinct contrast. (d) Two-dimensional birefringence map of the same sample area. The false-colour image encodes the orientation of the optical principal axis φ of the reflected light. Adjacent domains exhibit an approximately 90° rotation of the optical principal axis, consistent with a twinned structure.

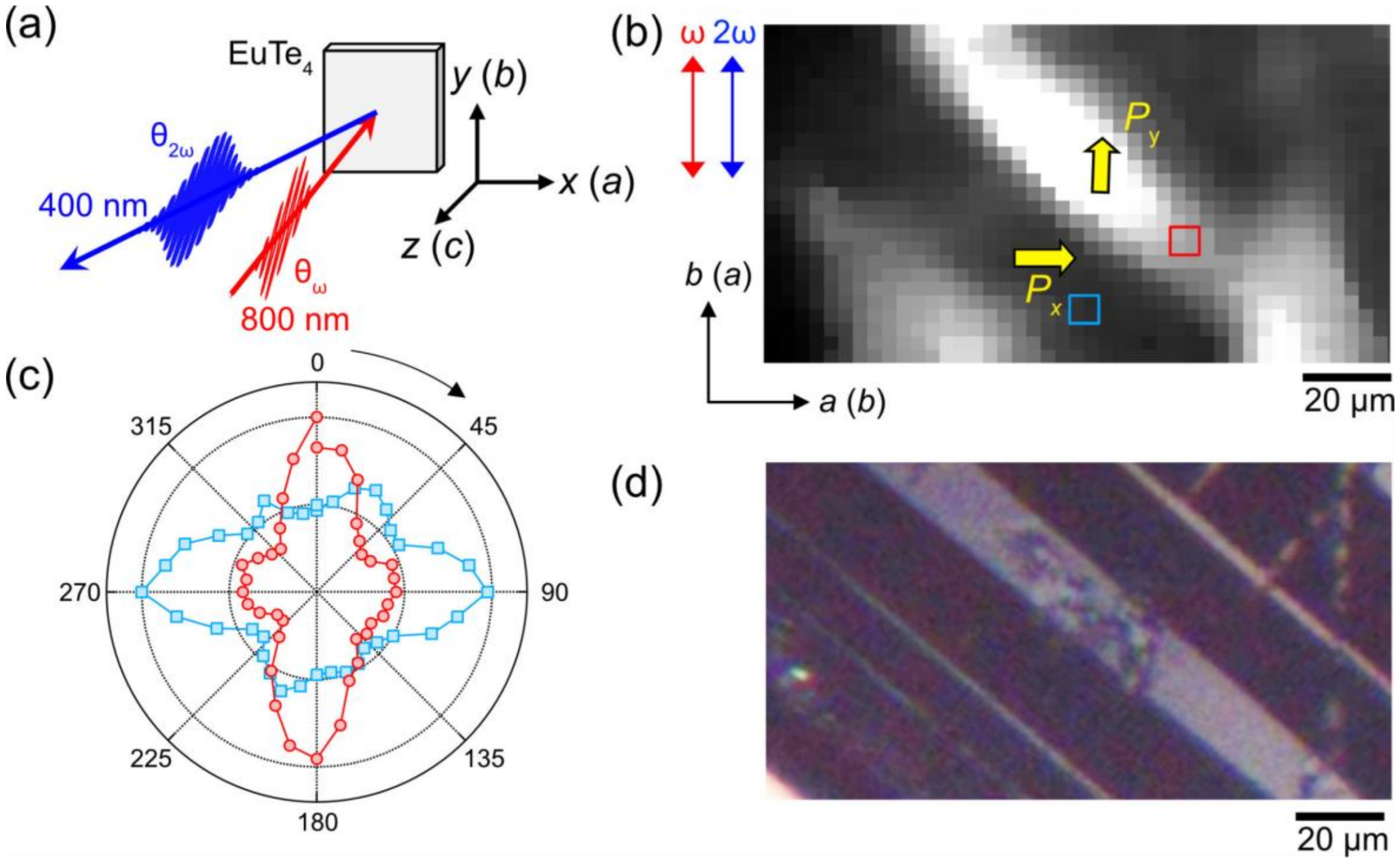

**Fig. 2 Optical second-harmonic generation (SHG) from a single crystal of $EuTe_4$.** (a) Schematic of the SHG measurement setup. The polarisation directions of the incident fundamental light (ω) and the emitted second harmonic light (2ω) are indicated by red and blue arrows, respectively. (b) Two-dimensional SHG intensity map acquired from the crystal surface. (c) Polarisation-angle dependence of the SHG intensity for two distinct domains in $EuTe_4$. The red circles and blue squares correspond to the regions marked by the red and blue boxes in Fig. 2b, respectively; the inferred in-plane polarisation directions for each domain are indicated by the yellow arrows in Fig. 2b. (d) Polarised-optical micrograph of the identical area on the same crystal shown in (b).

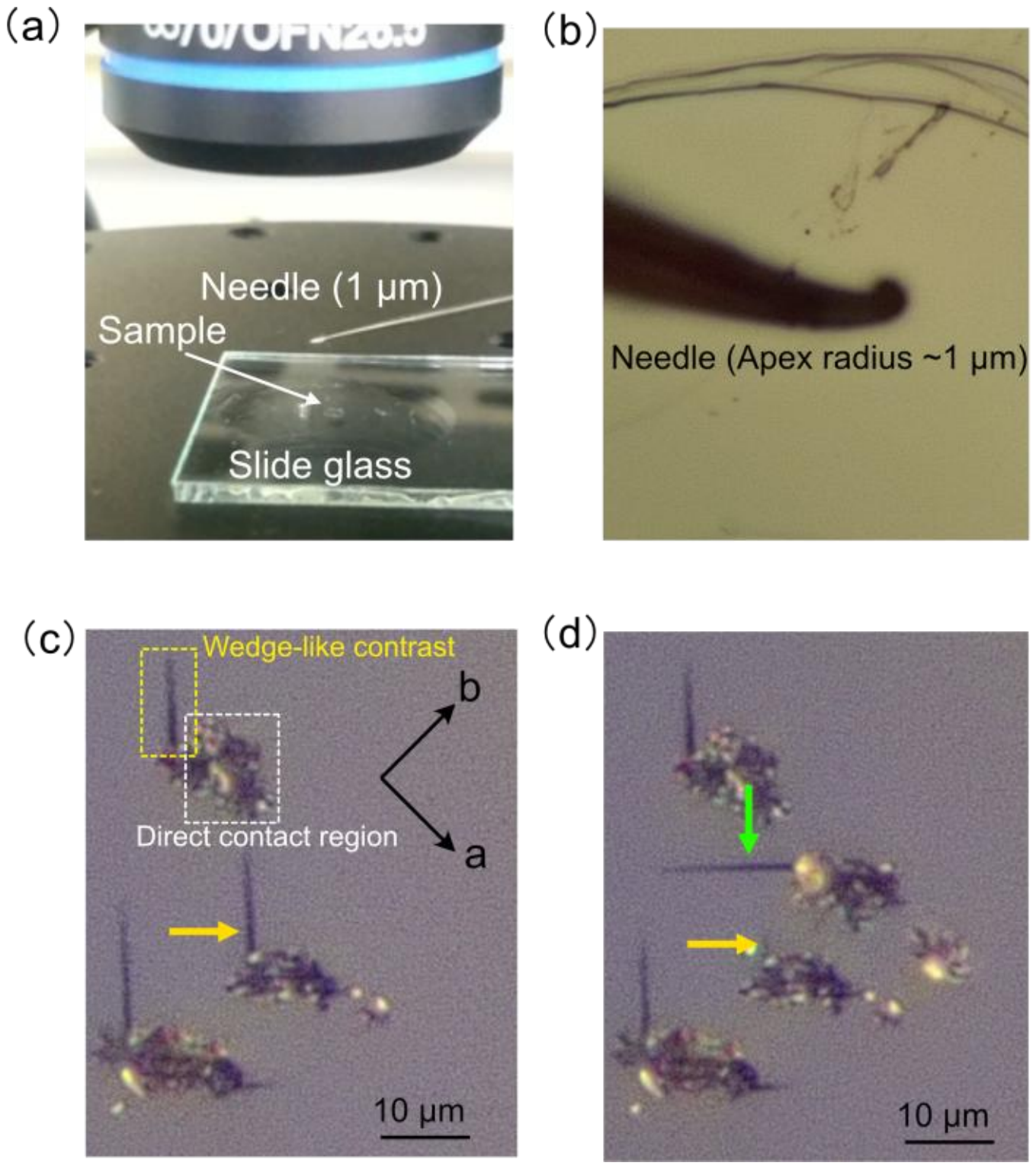

**Fig. 3 Indentation-induced domain reconfiguration in $EuTe_4$.** (a) Photograph of the experimental setup in which a needle with a 1 μm apex radius is pressed onto the sample while polarised optical micrographs are recorded. (b) Microscope view of the needle above the sample surface. (c) Polarised optical micrograph acquired after pressing the needle against the sample. (d) Image after applying local strain: the domain contrast highlighted in yellow in (c) is suppressed, and a new domain-like contrast feature appears at the position indicated by the green arrow.

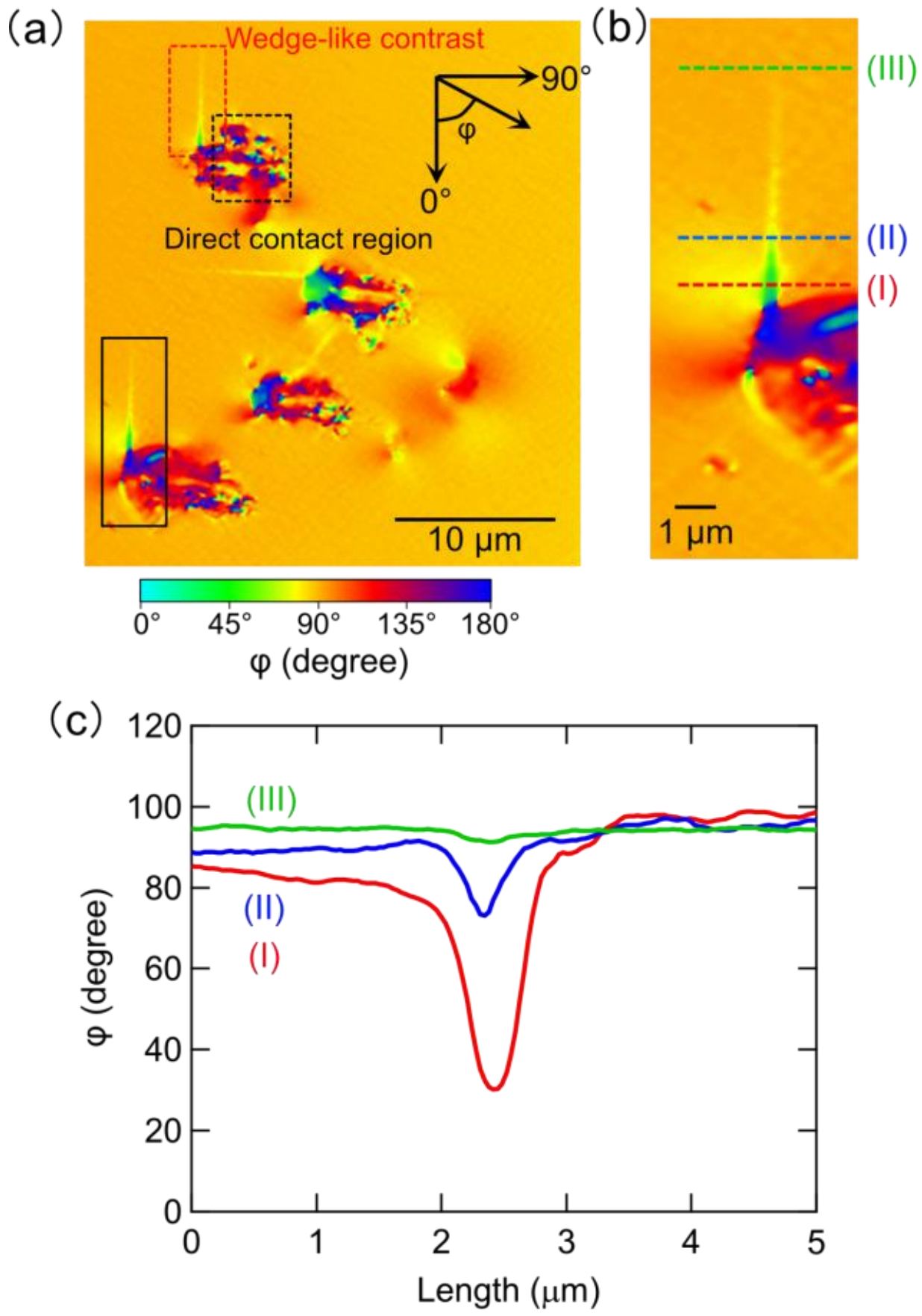

**Fig. 4 Birefringence measurement.** (a) Two-dimensional birefringence map after indentation. φ denotes the orientation angle of the optical principal axis. (b) Magnified view of the area outlined by the black box in (a). (c) Spatial variation of φ along the three line cuts (I–III) marked by dashed lines in (b).

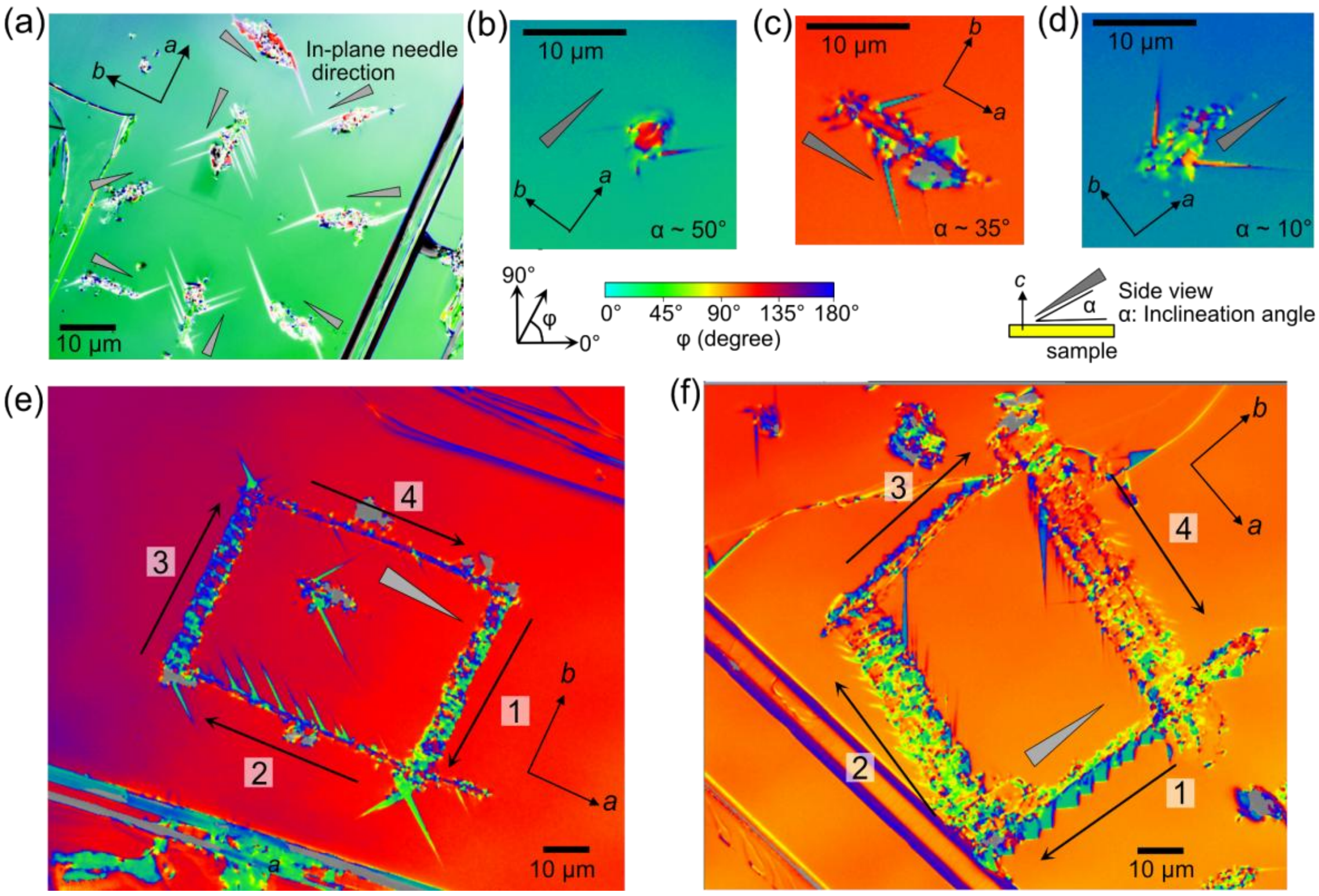

**Fig. 5 Dependence of mechanically induced domain reconfiguration on needle geometry and scanning direction.** (a) Birefringence image showing mechanically modified regions produced with different needle orientations relative to the crystallographic *a* and *b* axes. (b–d) Birefringence images obtained after local indentation with different inclination angles α between the needle and the sample surface: α ~ 50° (b), α ~ 35° (c), and α ~ 10° (d). The schematic illustrates that α denotes the out-of-plane inclination angle between the needle and the sample surface. (e,f) Birefringence images obtained after sequential mechanical scanning with the needle oriented approximately parallel to the crystallographic *a* axis (e) or *b* axis (f). Black arrows indicate the scanning direction, and the numbers 1–4 indicate the scanning order. Gray triangles indicate the needle-tip directions. In all panels, the colour scale represents the orientation angle φ of the optical principal axis, measured counterclockwise from the rightward horizontal direction in each image.

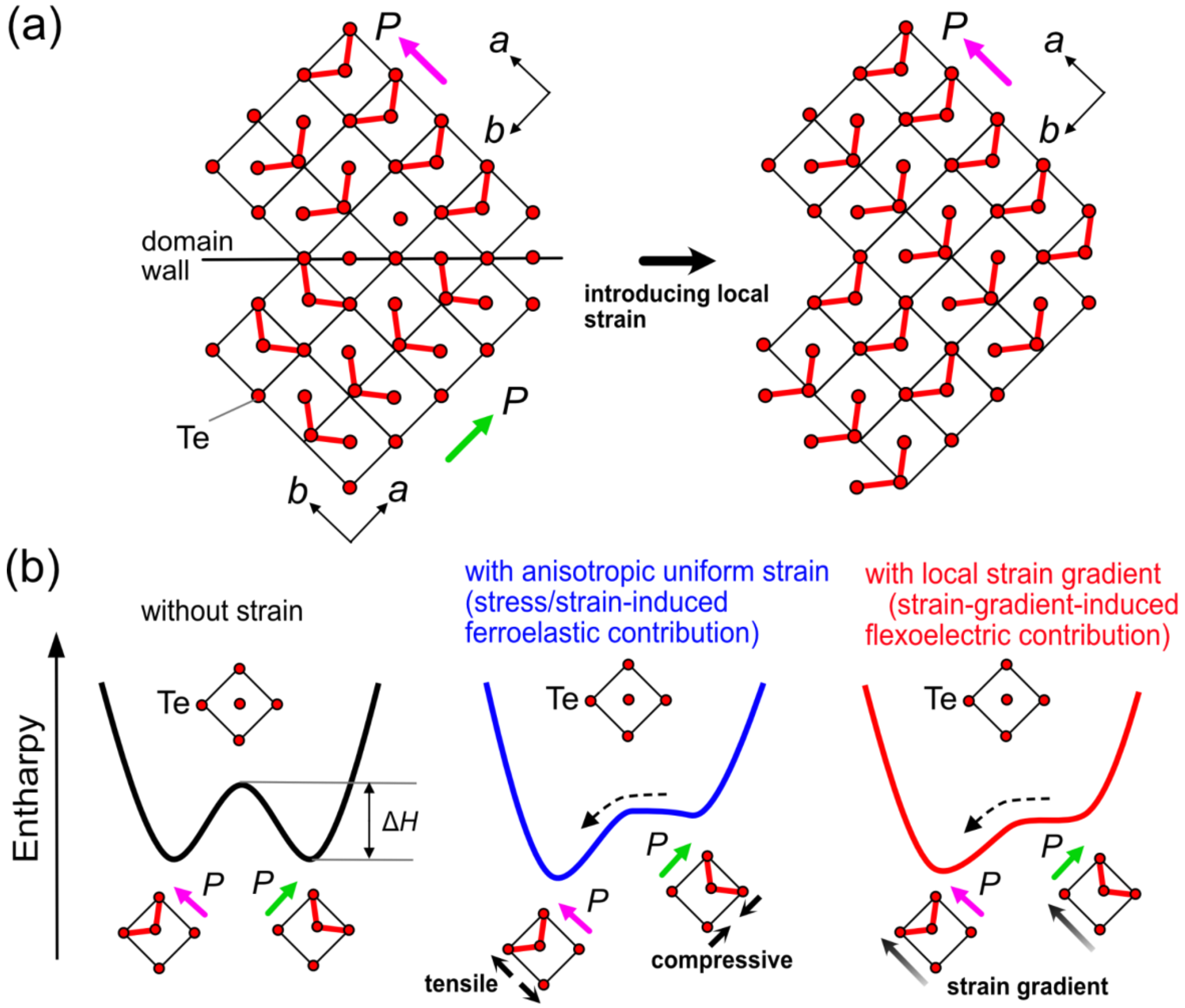

**Fig. 6 Polarisation and domain topology of the CDW state.** (a) Schematic of domain-wall formation in the CDW state, arising from Te-atom displacements within the Te layer. (b) Schematic enthalpy landscape illustrating the relative stability between the non-CDW and CDW states, $\Delta H$, in the absence of strain, under uniform anisotropic strain, and under a local strain-gradient-induced flexoelectric bias.

**Table 1. Calculated enthalpy difference ($\Delta H \equiv H_{CDW} - H_{non}$) between the non-CDW phase and the primitive CDW phase with and without applied strain, assuming orthorhombic symmetry for both phases.**

| | $a$-axis stress (kbar) | $b$-axis stress (kbar) | $\Delta H$ (eV/f.u.) |
|---|---|---|---|
| Strain (stress) free | 0 | 0 | -0.0164 |
| $a$-axis compression | -2.0 | 2.0 | -0.0145 |
| $a$-axis tension | 2.0 | -2.0 | -0.0175 |

## REFERENCES

1. Scott, J. F. Applications of modern ferroelectrics. *Science* **315**, 954–959 (2007).
2. Martin, L. W. & Rappe, A. M. Thin-film ferroelectric materials and their applications. *Nat. Rev. Mater.* **2**, 16087 (2017).
3. Dawber, M., Rabe, K. M. & Scott, J. F. Physics of thin-film ferroelectric oxides. *Rev. Mod. Phys.* **77**, 1083–1130 (2005).
4. Merz, W. J. Domain formation and domain wall motions in ferroelectric $BaTiO_3$ Single crystals. *Phys. Rev.* **95**, 690–698 (1954).
5. Setter, N. *et al.* Ferroelectric thin films: Review of materials, properties, and applications. *J. Appl. Phys.* **100**, 051606 (2006).
6. Park, S. M. *et al.* Selective control of multiple ferroelectric switching pathways using a trailing flexoelectric field. *Nat. Nanotechnol.* **13**, 366–370 (2018).
7. Kogan, S. M. Piezoelectric effect during inhomogeneous deformation and acoustic scattering of carriers in crystals. *Sov. Phys. Solid State* **5**, 2069–2070 (1964).
8. Catalan, G. *et al.* Flexoelectric rotation of polarization in ferroelectric thin films. *Nat. Mater.* **10**, 963–967 (2011).
9. Lu, H. *et al.* Mechanical writing of ferroelectric polarization. *Science* **336**, 59–61 (2012).

10. Lu, H. *et al.* Nanodomain engineering in ferroelectric capacitors with graphene electrodes. *Nano Lett.* **16**, 6460–6466 (2016).

11. Lee, D., Yang, S. M., Yoon, J.-G. & Noh, T. W. Flexoelectric rectification of charge transport in strain-graded dielectrics. *Nano Lett.* **12**, 6436–6440 (2012).

12. Li, Y.-J. *et al.* Mechanical switching of nanoscale multiferroic phase boundaries. *Adv. Funct. Mater.* **25**, 3405–3413 (2015).

13. Lu, X. *et al.* Mechanical-force-induced non-local collective ferroelastic switching in epitaxial lead-titanate thin films. *Nat. Commun.* **10**, 3951 (2019).

14. Li, Q. *et al.* Ferroelastic nanodomain-mediated mechanical switching of ferroelectricity in thick epitaxial films. *Nano Lett.* **21**, 445–452 (2021).

15. Liu, H. *et al.* Reversible flexoelectric domain engineering at the nanoscale in van der Waals ferroelectrics. *Nat. Commun.* **15**, 4556 (2024).

16. Ming, W. *et al.* Flexoelectric engineering of van der Waals ferroelectric $CuInP_2S_6$. *Sci. Adv.* **8**, eabq1232 (2022).

17. Chang, K. *et al.* Discovery of robust in-plane ferroelectricity in atomic-thick SnTe. *Science* **353**, 274–278 (2016).

18. Shi, Y. *et al.* A ferroelectric-like structural transition in a metal. *Nat. Mater.* **12**, 1024–1027 (2013).

19. Fei, Z. *et al.* Ferroelectric switching of a two-dimensional metal. *Nature* **560**, 336–339 (2018).

20. Lv, B. Q. *et al.* Experimental Discovery of Weyl Semimetal TaAs. *Phys. Rev. X* **5**, 031013 (2015).

21. Jindal, A. *et al.* Coupled ferroelectricity and superconductivity in bilayer Td-$MoTe_2$. *Nature* **613**, 48–52 (2023).

22. Ishizaka, K. *et al.* Giant Rashba-type spin splitting in bulk BiTeI. *Nat. Mater.* **10**, 521–526 (2011).

23. Di Sante, D., Barone, P., Bertacco, R. & Picozzi, S. Electric control of the giant Rashba effect in bulk GeTe. *Adv. Mater.* **25**, 509–513 (2013).

24. Kolobov, A. V. *et al.* Ferroelectric switching in epitaxial GeTe films. *APL Mater.* **2**, 066101 (2014).

25. Zabalo, A. & Stengel, M. Switching a polar metal via strain gradients. *Phys. Rev. Lett.* **126**, 127601 (2021).

26. Peng, W. *et al.* Flexoelectric polarizing and control of a ferromagnetic metal. *Nat. Phys.* **20**, 450–455 (2024).

27. Lun, Y. *et al.* Polarization vortices in a ferromagnetic metal via twistronics. *Nat. Mater.* (2026)

28. Chu, Y.-H. *et al.* Electric-field control of local ferromagnetism using a magnetoelectric multiferroic. *Nat. Mater.* **7**, 478–482 (2008).

29. Livesey, K. L. Strain-mediated magnetoelectric coupling in magnetostrictive/piezoelectric heterostructures and resulting high-frequency effects. *Phys. Rev. B* **83**, 224420 (2011).

30. Chen, A. T. & Zhao, Y. G. Research Update: Electrical manipulation of magnetism through strain-mediated magnetoelectric coupling in multiferroic heterostructures. *APL Mater.* **4**, 032303 (2016).

31. Sakano, M. *et al.* Observation of spin-polarized bands and domain-dependent Fermi arcs in polar Weyl semimetal $MoTe_2$. *Phys. Rev. B.* **95**, 121101(R) (2017).

32. Berger, A. N. *et al.* Temperature-driven topological transition in 1T'-$MoTe_2$. *npj Quantum Mater.* **3**, 2 (2018).

33. Biswas, K. *et al.* High-performance bulk thermoelectrics with all-scale hierarchical architectures. *Nature* **489**, 414–418 (2012).

34. Zhang, Q. Q. *et al.* Thermal hysteretic behavior and negative magnetoresistance in the charge density wave material $EuTe_4$. *Phys. Rev. B.* **107**, 115141 (2023).

35. Zhang, C. *et al.* Angle-resolved photoemission spectroscopy study of charge density wave

order in the layered semiconductor $EuTe_4$. *Phys. Rev. B.* **106**, L201108 (2022).

36. Lv, B. Q. *et al.* Unconventional hysteretic transition in a charge density wave. *Phys. Rev. Lett.* **128**, 036401 (2022).
37. Elius, I. B. *et al.* Observation of momentum dependent charge density wave gap in $EuTe_4$. *Phys. Rev. B.* **113**, 235125 (2026).
38. Takahashi, H., Yoshida, K., Nakano, A. & Ishiwata, S. Enhanced thermopower in a magnetic semiconductor $EuTe_4$ with multiple charge-density-wave instabilities. *PRX Energy* **4**, 033009 (2025).
39. Wu, D. *et al.* Layered semiconductor $EuTe_4$ with charge density wave order in square tellurium sheets. *Phys. Rev. Mater.* **3**, 024002 (2019).
40. Rathore, R. *et al.* Evolution of static charge density wave order, amplitude mode dynamics, and suppression of Kohn anomalies at the hysteretic transition in $EuTe_4$. *Phys. Rev. B.* **107**, 024101 (2023).
41. Lv, B. Q. *et al.* Large moiré superstructure of stacked incommensurate charge density waves. *Nat. Mater.* **25**, 420–426 (2025).
42. Oh, K. H. *et al.* Dynamics of a jointly commensurate moiré charge density wave. *Nat. Commun.* 1–9 (2026).
43. Matsubara, M. *et al.* Multiferroics. Magnetoelectric domain control in multiferroic $TbMnO_3$. *Science* **348**, 1112–1115 (2015).
44. Kondo, M. *et al.* Tunable spin-valley coupling in layered polar Dirac metals. *Commun. Mater.* **2**, 49 (2021).
45. Shi, C. *et al.* Domain-dependent strain and stacking in two-dimensional van der Waals ferroelectrics. *Nat. Commun.* **14**, 7168 (2023).
46. Straquadine, J. A. W., Ikeda, M. S. & Fisher, I. R. Evidence for realignment of the charge density wave state in $ErTe_3$ and $TmTe_3$ under uniaxial stress via elastocaloric and elastoresistivity measurements. *Phys. Rev. X.* **12**, 021046 (2022).

47. Lun, Y. *et al.* Ultralow tip-force driven sizable-area domain manipulation through transverse flexoelectricity. *Adv. Mater.* **35**, e2302320 (2023).

48. Wang, B. *et al.* Mechanically induced ferroelectric switching in $BaTiO_3$ thin films. *Acta Mater.* **193**, 151–162 (2020).

49. Jiang, N. *et al.* Zero-field superconducting diode effect induced by magnetic flux in the van der Waals trigonal superconductor $PtBi_2$. *Phys. Rev. B.* **112**, 235313 (2025).

50. Perdew, J. P., Burke, K. & Ernzerhof, M. Generalized gradient approximation made simple. *Phys. Rev. Lett.* **77**, 3865–3868 (1996).

51. Kresse, G. & Hafner, J. Ab initio molecular dynamics for liquid metals. *Phys. Rev. B* **47**, 558–561 (1993).

52. Kresse, G. & Hafner, J. Ab initio molecular-dynamics simulation of the liquid-metal-amorphous-semiconductor transition in germanium. *Phys. Rev. B* **49**, 14251–14269 (1994).

53. Kresse, G. & Furthmüller, J. Efficiency of ab-initio total energy calculations for metals and semiconductors using a plane-wave basis set. *Comput. Mater. Sci.* **6**, 15–50 (1996).

54. Kresse, G. & Furthmüller, J. Efficient iterative schemes for ab initio total-energy calculations using a plane-wave basis set. *Phys. Rev. B* **54**, 11169–11186 (1996).

55. Kresse, G. & Joubert, D. From ultrasoft pseudopotentials to the projector augmented-wave method. *Phys. Rev. B* **59**, 1758–1775 (1999).

56. Blöchl, P. E. Projector augmented-wave method. *Phys. Rev. B* **50**, 17953–17979 (1994).

57. Pathak, A., Gupta, M. K., Mittal, R. & Bansal, D. Orbital- and atom-dependent linear dispersion across the Fermi level induces charge density wave instability in $EuTe_4$. *Phys. Rev. B.* **105**, 035120 (2022).

58. Momma, K. & Izumi, F. VESTA 3 for three-dimensional visualization of crystal, volumetric and morphology data. *J. Appl. Crystallogr.* **44**, 1272–1276 (2011).